\documentclass[11pt]{article}
\usepackage{amsfonts,amssymb,amsmath}

\numberwithin{equation}{section} 

\def\LRA{\mathop{-\!\!\!-\!\!\!
\longrightarrow}\nolimits}
\newcommand{\calkaob}[2]{\int_{\mathbb{R}^{3}}\mathrm{d}^{3}
\wektor{#1}#2}
\newcommand{\calkapow}[2]{\oint_{\mathrm{4\pi}}\mathrm{d}^{2}
{\hat{\wektor{#1}}}#2}
\newcommand{\Amplituda}[0]{A(\wektor{k}_{f}
\leftarrow\wektor{k}_{i})}
\newcommand{\mac}[1]{\mathsf{#1}}
\newcommand{\wektor}[1]{\boldsymbol{#1}}

\newcommand{\braket}[2]{\big<#1\big|#2\big>}
\begin{document}

\title{Non-relativistic quantum scattering from non-local separable
potentials: the eigenchannel approach}
\author{Remigiusz Augusiak\footnote{email: remik@mif.pg.gda.pl}\\
\small{Department of Theoretical Physics and Mathematical Methods},\\
\small{Faculty of Applied Physics and Mathematics},\\
\small{Gda\'nsk University of Technology}, \\
\small{Narutowicza 11/12, PL 80--952 Gda\'nsk, Poland}}

\date{}

\maketitle

\abstract{A recently formulated version of the eigenchannel method
[R. Szmytkowski, Ann. Phys. (N.Y.) {\bf 311}, 503 (2004)] is
applied to quantum scattering of Schr\"odinger particles from
non-local separable potentials. Eigenchannel vectors and negative
cotangents of eigenphase-shifts are introduced as eigensolutions
to some weighted matrix spectral problem, without a necessity of
prior construction and diagonalization of the scattering matrix.
Explicit expressions for the scattering amplitude, the total cross
section in terms of the eigenchannel vectors and the
eigenphase-shifts are derived. An illustrative example is
provided.}

\section{Introduction}
\label{wstep}

\noindent\indent Recently, Szmytkowski \cite{Szmytkowski} (see
also \cite{Szmytkowski2}), inspired by works of Garbacz
\cite{Garbacz1} and Harrington and Mautz \cite{Harrington1} has
presented a general formulation of the eigenchannel method for
quantum scattering from Hermitian short-range potentials. This
formulation, exploiting the formalism of integral equations, is
alternative to that proposed in 1960's by Danos and Greiner
\cite{Danos}, who based on wave equations written in differential
forms.

While various theoretical aspects of the new approach to the
eigenchannel method have been considered in detail in Ref.
\cite{Szmytkowski}, no explicit illustrative applications of the
formalism have been provided therein. It is the purpose of the
present paper to fill in this gap by applying the formalism to
non-relativistic scattering from non-local separable potentials.

Separable potentials have proved to be useful in many branches of
physics such as atomic \cite{Ponomarenko}, molecular
\cite{Molecular} and solid state \cite{Prunele1} physics. Still,
the most extensive use of the separable potentials seems to be
made in nuclear physics for describing nucleon-nucleon
interactions \cite{NN}. The utility of separable potentials stems
from two facts. Firstly, the Lippman-Schwinger \cite{LippSchw}
equation with a separable potential may be solved by employing
algebraic methods. Secondly, methods are known which allow one to
construct separable approximations to arbitrary local potentials
\cite{metody}. Here, the crucial role plays the method proposed by
Ernst, Shakin and Thaler \cite{EST} (see also \cite{Zast_EST}).

The arrangement of the paper is as follows. Section
\ref{rozdzial1} contains some basic notions of quantum theory of
potential scattering. In Section \ref{rozdzial2} we focus on
scattering from separable potentials. In Section 4 we define
\textit{eigenchannel vectors} as eigenvectors of some weighted
spectral problem and investigate some of their properties. We
introduce also eigenphaseshifts using eigenvalues of the same
spectral problem. Then, we derive formulas for a scattering
amplitude in terms of eigenchannel vectors and calculate a total
cross section and an averaged total cross section. Finally, in
Section 5 we consider an illustrative example. The paper ends with
two appendices.
\section{Non-relativistic quantum scattering from non-local potentials}
\label{rozdzial1} \noindent\indent Consider a Schr\"odinger
particle described by the monochromatic plane wave
\begin{equation}\label{PlaneWave}
\phi_{i}(\wektor{r})\equiv\braket{\wektor{r}}{\wektor{k}_{i}}
=e^{\mathrm{i}\wektor{k}_{i}\cdot\wektor{r}}
\end{equation}
with $\hbar\wektor{k}_{i}$ denoting its initial momentum,
scattered elastically, i.e. $|\wektor{k}_{i}|=|\wektor{k}_{f}|=k$
($\hbar\wektor{k}_{f}$ is a momentum of the scattered particle) by
a non-local Hermitian potential with a kernel
$V(\wektor{r},\wektor{r}')$.

It was shown by Lippmann and Schwinger \cite{LippSchw} that such
processes can be described by the following integral equation:
\begin{equation}\label{2.1}
\psi(\wektor{r})= \phi_{i}(\wektor{r})-\calkaob{r}'\calkaob{r}''\:
G(E,\wektor{r},\wektor{r}') V(\wektor{r}',\wektor{r}'')
\psi(\wektor{r}''),
\end{equation}
where $G(E,\wektor{r},\wektor{r}')$ is the well-known
free-particle outgoing Green function and is of the form
\begin{equation}\label{2.2}
G(E,\wektor{r},\wektor{r}') =\frac{m}{2\pi\hbar^{2}}
\frac{\mathrm{e}^{\mathrm{i}k|\wektor{r}-\wektor{r}'|}}
{|\wektor{r}-\wektor{r}'|}.
\end{equation}
As usual, during considerations concerning scattering processes we
tend to find asymptotical behavior of the wave function
$\psi(\wektor{r})$. To aim this we firstly need to find
asymptotical form of Green function (\ref{2.2}) for $r\to\infty$.
After straightforward movements we have:
\begin{equation}\label{2.3}
G(E,\wektor{r},\wektor{r}') \stackrel{r\to\infty}{\LRA}
\frac{m}{2\pi\hbar^{2}}\frac{\mathrm{e}^{\mathrm{i}kr}}{r}
\mathrm{e}^{-\mathrm{i}\wektor{k}_{f}\cdot\wektor{r}'}, \qquad
\wektor{k}_{f}=k\frac{\wektor{r}}{r}.
\end{equation}
and therefore
\begin{equation}\label{2.4}
\psi(\wektor{r})\stackrel{r\to\infty}
{\LRA}\mathrm{e}^{\mathrm{i}\boldsymbol{k}_{i}\cdot\wektor{r}}
+\Amplituda\frac{\mathrm{e}^{\mathrm{i}kr}}{r}.
\end{equation}
The quantity $\Amplituda$ appearing above is called scattering
amplitude and is of the form
\begin{equation}\label{2.5}
\Amplituda=-\frac{m}{2\pi\hbar^{2}}
\calkaob{r}'\calkaob{r}''\:\mathrm{e}^
{-\mathrm{i}\wektor{k}_{f}\cdot\wektor{r}'}
V(\wektor{r}',\boldsymbol{r}'')\psi(\wektor{r}'').
\end{equation}
In terms of the scattering amplitude the differential cross
section for scattering from the direction $\wektor{k}_{i}$ into
the direction $\wektor{k}_{f}$ is
\begin{equation}\label{2.6}
\frac{\mathrm{d}\sigma}{\mathrm{d}\Omega_{f}}
(\wektor{k}_{f}\leftarrow\wektor{k}_{i})= |\Amplituda|^{2}.
\end{equation}
Moreover, one defines the total cross section in the following way
\begin{equation}\label{2.7}
\sigma(\wektor{k}_{i})= \calkapow{k}_{f}\:|\Amplituda|^{2}.
\end{equation}
Then averaging the total cross section by all directions of
incidence $\wektor{k}_{i}$, one gets the so-called averaged total
cross section:
\begin{eqnarray}\label{2.8}
\sigma_{t}(E)=
\frac{1}{4\pi}\calkapow{k}_{i}\calkapow{k}_{f}\:|\Amplituda|^{2}.
\end{eqnarray}
\section{The special class of non-local separable potentials}
\label{rozdzial2} In this section we focus on the special class of
Hermitian separable potentials given by the following kernel:
\begin{equation}\label{3.2}
V(\wektor{r},\wektor{r}')=\sum_{\mu}\omega_{\mu}
v_{\mu}(\wektor{r})v_{\mu}^{*}(\wektor{r}') \qquad
\left(\forall\mu :\,\,\omega_{\mu}\in\mathbb{R}\setminus\{0\}
\right),
\end{equation}
where $\mu$ may, in general, denote an arbitrary finite set of
countable indices $\mu_{i}$, i.e.,
$\mu=\{\mu_{1},\ldots,\mu_{k}\}$ and asterisk stands for the
complex conjugation.

Application of Eq. (\ref{3.2}) to Eq. (\ref{2.1}) leads us to the
Lippmann--Schwinger equation for the separable potentials:
\begin{equation}\label{3.3}
\psi(\wektor{r})= \phi_{i}(\wektor{r})-\sum_{\mu}\omega_{\mu}
\calkaob{r}'\:G(E,\wektor{r},\wektor{r}')
v_{\mu}(\wektor{r}')\calkaob{r}''\:
v_{\mu}^{*}(\wektor{r}'')\psi(\wektor{r}'').
\end{equation}
Similarly, substitution of Eq. (\ref{3.2}) to Eq. (\ref{2.5})
gives us the scattering amplitude of the form
\begin{equation}\label{3.4}
\Amplituda=-\frac{m}{2\pi\hbar^{2}} \sum_{\mu}\omega_{\mu}
\calkaob{r}'\:
\mathrm{e}^{-\mathrm{i}\wektor{k}_{f}\cdot\wektor{r}'}v_{\mu}(\wektor{r}')
\calkaob{r}''\:v_{\mu}^{*}(\wektor{r}'')\psi(\wektor{r}'').
\end{equation}
For the sake of convenience, henceforth we shall use the Dirac
notation. Then, the above formula reads
\begin{equation}\label{3.5}
\Amplituda=-\frac{m}{2\pi\hbar^{2}}
\sum_{\mu}\big<\wektor{k}_{f}\big|v_{\mu}\big> \omega_{\mu}
\big<v_{\mu}\big|\psi\big>.
\end{equation}
The fact that $\mu$ is an element of an arbitrary countable set,
implies that we can put all scalar products
$\braket{v_{\mu}}{\psi}$ into a sequence. Therefore, for further
simplicity we can employ the following notations:
\begin{equation}\label{3.6}
\braket{\wektor{v}}{\varphi}= \left(
\begin{array}{c}
\braket{v_{1}}{\varphi}\\*[0.5ex]
\braket{v_{2}}{\varphi}\\
\vdots
\end{array}
\right), \qquad \braket{\varphi}{\wektor{v}}=
\braket{\wektor{v}}{\varphi}^{\dag}= \left(
\braket{\varphi}{v_{1}}\; \braket{\varphi}{v_{2}}\;\dots \right)
\end{equation}
and
\begin{equation}\label{3.7}
\mac{\Omega}=\mathrm{diag}[\omega_{1},\omega_{2},\dots],
\end{equation}
where the dagger denotes the matrix Hermitian conjugation. Note
that from Eqs. (\ref{3.2}) and (\ref{3.7}) it is evident that the
matrix $\mac{\Omega}$ is invertible. We keep this fact for
purposes of further considerations. In the light of Eqs.
(\ref{3.6}) and (\ref{3.7}) we may rewrite the scattering
amplitude (\ref{3.5}) in the following form:
\begin{equation}\label{3.8}
\Amplituda=-\frac{m}{2\pi\hbar^{2}}
\braket{\wektor{k}_{f}}{\wektor{v}}\mac{\Omega}
\braket{\wektor{v}}{\psi}.
\end{equation}
In the last step we need to calculate the scalar product
$\braket{\wektor{v}}{\psi}$. To this end, we transform
Lippmann--Schwinger equation for the separable
potentials~(\ref{3.3}) to a set of linear algebraic equations.
Hence, after some elementary movements, we have
\begin{equation}\label{3.9}
\sum_{\mu} \left[ \delta_{\nu\mu}+ \mac{G}_{\nu\mu}\omega_{\mu}
\right] \braket{v_{\mu}}{\psi}= \braket{v_{\nu}}{\wektor{k}_{i}},
\end{equation}
where
\begin{equation}\label{elG}
\mac{G}_{\nu\mu}=\calkaob{r}\calkaob{r}'\:v_{\nu}^{*}(\wektor{r})
G(E,\wektor{r},\wektor{r}')v_{\mu}(\wektor{r}').
\end{equation}
Finally, application of Eqs. (\ref{3.6}) and (\ref{3.7}) to Eq.
(\ref{3.9}) yields
\begin{equation}\label{3.10}
(\mac{I}+\mac{G}\mac{\Omega})
\braket{\wektor{v}}{\psi}=\braket{\wektor{v}}{\wektor{k}_{i}}\quad
\Rightarrow\quad \braket{\wektor{v}}{\psi}=
(\mac{I}+\mac{G}\mac{\Omega})^{-1}
\braket{\wektor{v}}{\wektor{k}_{i}},
\end{equation}
where $\mac{I}$ stands for the identity matrix and $\mac{G}$
denotes a matrix with the elements $\mac{G}_{\nu\mu}$.
Substitution of Eq. (\ref{3.10}) to Eq. (\ref{3.8}) gives the
expression for the scattering amplitude
\begin{equation}\label{3.11}
\Amplituda=-\frac{m}{2\pi\hbar^{2}}
\braket{\wektor{k}_{f}}{\wektor{v}}\mac{\Omega} \left(
\mac{I}+\mac{G}\,\mac{\Omega} \right)^{-1}
\braket{\wektor{v}}{\wektor{k}_{i}}.
\end{equation}
Since $(\mac{M}\mac{N})^{-1}=\mac{N}^{-1}\mac{M}^{-1}$ for all
invertible matrices $\mac{M}$ and $\mac{N}$, we can conclude that
\begin{equation}\label{3.12}
\Amplituda=-\frac{m}{2\pi\hbar^{2}}
\braket{\wektor{k}_{f}}{\wektor{v}} \left(
\mac{\Omega}^{-1}+\mac{G} \right)^{-1}
\braket{\wektor{v}}{\wektor{k}_{i}}.
\end{equation}
\section{The eigenchannel method}
\label{rozdzial3} \noindent\indent Here, we turn to the
formulation of the eigenchannel method proposed by Szmytkowski
\cite{Szmytkowski}. This author has shown that the eigenchannels
provide a powerful mathematical tool to the quantum theory of
scattering. Moreover, in this approach we \textit{do not need} to
construct the scattering matrix to obtain the formula for an
averaged total cross section.

In the first step we rewrite the matrix
$\mac{\Omega}^{-1}+\mac{G}$ as a sum of its Hermitian and
non--Hermitian parts. Hence, we have
\begin{equation}\label{4.1}
\mac{\Omega}^{-1}+\mac{G}=\mac{A}+\mathrm{i}\mac{B},
\end{equation}
where the matrices $\mac{A}$ and $\mac{B}$ are defined by the
relations:
\begin{equation}\label{4.2}
\mac{A}=\frac{1}{2} \left[ \mac{\Omega}^{-1}+\mac{G}+
(\mac{\Omega}^{-1}+\mac{G})^{\dag} \right]=
\mac{\Omega}^{-1}+\frac{1}{2} (\mac{G}+\mac{G}^{\dag})
\end{equation}
and
\begin{equation}\label{4.3}
\mac{B}=\frac{1}{2\mathrm{i}} \left[ \mac{\Omega}^{-1}+\mac{G}-
(\mac{\Omega}^{-1}+\mac{G})^{\dag} \right]= \frac{1}{2\mathrm{i}}
(\mac{G}-\mac{G}^{\dag}),
\end{equation}
respectively. From the definitions (\ref{4.2}) and (\ref{4.3}) it
is evident that both matrices $\mac{A}$ and $\mac{B}$ are
Hermitian, i.e., $\mac{A}=\mac{A}^{\dag}$ and
$\mac{B}=\mac{B}^{\dag}$. Moreover, straightforward calculations
with the aid of Eqs. (\ref{2.2}) and (\ref{elG}) allow us to
express their elements as
\begin{equation}\label{elA}
[\mac{A}]_{\nu\mu}= \frac{1}{\omega}_{\nu}\delta_{\nu\mu}+
\frac{m}{2\pi\hbar^{2}}\calkaob{r}\calkaob{r}'\:v_{\nu}^{*}(\wektor{r})
\frac{\cos(k|\wektor{r}-\wektor{r}'|)}{|\wektor{r}-\wektor{r}'|}
v_{\mu}(\wektor{r}'),
\end{equation}
\begin{eqnarray}\label{elB}
[\mac{B}]_{\nu\mu}=
\frac{m}{2\pi\hbar^{2}}\calkaob{r}\calkaob{r}'\:v_{\nu}^{*}(\wektor{r})
\frac{\sin(k|\wektor{r}-\wektor{r}'|)}{|\wektor{r}-\wektor{r}'|}
v_{\mu}(\wektor{r}').
\end{eqnarray}
Following \cite{Szmytkowski} let us consider the following
weighted spectral problem
\begin{equation}\label{4.6}
\mac{A}X_{\gamma}(E)=\lambda_{\gamma}(E)\mac{B}X_{\gamma}(E),
\end{equation}
where $X_{\gamma}(E)$ and $\lambda_{\gamma}(E)$ are, respectively,
its eigenvectors and eigenvalues. Throughout the rest of the
present paper, the eigenvectors $X_{\gamma}(E)$ will be called
\textit{eigenchannel vectors}. They constitute a representation of
eigenchannels, defined in \cite{Szmytkowski} as state vectors, in
subspace spanned by the potential functions $v_{\mu}(\wektor{r})$.
In fact, the knowledge of eigenvectors $X_{\gamma}(E)$ allows us,
by some elementary steps, to construct the eigenchannels.

By virtue of the fact that both $\mac{A}$ and $\mac{B}$ are
Hermitian with the aid of positive semidefiniteness of the matrix
$\mac{B}$ (for a proof, cf. Appendix \ref{appB}), we infer that
$\lambda_{\gamma}^{*}(E)=\lambda_{\gamma}(E)$ for all $\gamma$.
Moreover, it is easy to show that eigenvectors $X_{\gamma}(E)$
satisfy the weighted orthogonality relation
\begin{equation}\label{4.7}
X_{\gamma'}^{\dag}(E)\mac{B}X_{\gamma}(E)=0 \qquad (\gamma'\ne
\gamma).
\end{equation}
Hence, it is natural to assume the following normalization:
\begin{equation}\label{4.8}
X_{\gamma'}^{\dag}(E)\mac{B}X_{\gamma}(E)=\delta_{\gamma'\gamma},
\end{equation}
or, using the matrices $\mac{A}$ and $\mac{\Omega}^{-1}+\mac{G}$:
\begin{equation}\label{4.9}
X_{\gamma'}^{\dag}(E)\mac{A}X_{\gamma}(E)=
\lambda_{\gamma}(E)\delta_{\gamma'\gamma},\qquad
X_{\gamma'}^{\dag}(E)(\mac{\Omega}^{-1}+\mac{G})X_{\gamma}(E)=
[\mathrm{i}+\lambda_{\gamma}(E)]\delta_{\gamma'\gamma}.
\end{equation}
Since the eigenvectors $X_{\gamma}(E)$ are solutions of the
Hermitian eigenproblem, they should obey the following closure
relations:
\begin{equation}\label{4.10}
\sum_{\gamma}X_{\gamma}(E)X_{\gamma}^{\dag}(E)\mac{B}=\mac{I},
\qquad \sum_{\gamma}\lambda_{\gamma}^{-1}(E)
X_{\gamma}(E)X_{\gamma}^{\dag}(E)\mac{A}=\mac{I},
\end{equation}
and
\begin{equation}\label{4.11}
\sum_{\gamma}
\frac{1}{\mathrm{i}+\lambda_{\gamma}(E)}X_{\gamma}(E)X_{\gamma}^{\dag}(E)
(\mac{\Omega}^{-1}+\mac{G})=\mac{I}.
\end{equation}
Henceforth, we shall be assuming that the potential (\ref{3.2}) is
such that the above relations are satisfied. Therefore, it is
possible to express the matrix $(\mac{\Omega}^{-1}+\mac{G})^{-1}$
in terms of the eigenchannel vectors $X_{\gamma}(E)$. Indeed,
using Eq. (\ref{4.11}) we have
\begin{equation}\label{4.12}
(\mac{\Omega}^{-1}+\mac{G})^{-1}=\sum_{\gamma}
\frac{1}{\mathrm{i}+\lambda_{\gamma}(E)}X_{\gamma}(E)X_{\gamma}^{\dag}(E).
\end{equation}
Application of Eq. (\ref{4.12}) to Eq. (\ref{3.12}) yields
\begin{eqnarray}\label{4.13}
\Amplituda= -\frac{m}{2\pi\hbar^{2}}\sum_{\gamma}
\frac{1}{\mathrm{i}+\lambda_{\gamma}(E)}
\braket{\wektor{k}_{f}}{\wektor{v}}X_{\gamma}(E)
X_{\gamma}^{\dag}(E)\braket{\wektor{v}}{\wektor{k}_{i}}
\end{eqnarray}
or equivalently
\begin{equation}\label{4.14}
\Amplituda= -\frac{m}{2\pi\hbar^{2}}\sum_{\gamma}
\frac{1}{\mathrm{i}+\lambda_{\gamma}(E)}\sum_{\mu}
\braket{\wektor{k}_{f}}{v_{\mu}}X_{\gamma\mu}(E)\sum_{\nu}
X_{\gamma\nu}^{*}(E) \braket{v_{\nu}}{\wektor{k}_{i}} .
\end{equation}
Because of the very symmetrical form of the scattering amplitude,
it is useful to define the following functions:
\begin{equation}\label{Yfun}
\mathcal{Y}_{\gamma}(\wektor{k})=
\sqrt{\frac{mk}{8\pi^{2}\hbar^{2}}}\,\sum_{\mu}
\braket{\wektor{k}}{v_{\mu}}X_{\gamma\mu}(E)=\sqrt{\frac{mk}{8\pi^{2}\hbar^{2}}}\,
\braket{\wektor{k}}{\wektor{v}}X_{\gamma}(E),
\end{equation}
hereafter termed \textit{eigenchannel harmonics}. It follows from
their definition that they are orthonormal on the unit sphere (cf.
Appendix \ref{appA}), i.e.,
\begin{equation}\label{orthrel}
\calkapow{k}\;\mathcal{Y}_{\gamma'}^{*}(\wektor{k})\mathcal{Y}_{\gamma}(\wektor{k})=
\delta_{\gamma'\gamma}.
\end{equation}
After substitution of Eq. (\ref{Yfun}) to Eq. (\ref{4.14}), one
finds
\begin{eqnarray}\label{4.17}
\Amplituda=
-\frac{4\pi}{k}\sum_{\gamma}\frac{1}{\mathrm{i}+\lambda_{\gamma}(E)}
\mathcal{Y}_{\gamma}(\wektor{k}_{f})\mathcal{Y}_{\gamma}^{*}(\wektor{k}_{i}).
\end{eqnarray}
Further, it is convenient to express the eigenvalues
$\lambda_{\gamma}(E)$ in terms of so-called eigenphaseshifts
$\delta_{\gamma}(E)$ by the relation
\begin{equation}\label{4.18}
\lambda_{\gamma}(E)=-\cot\,\delta_{\gamma}(E),
\end{equation}
which after application to the scattering amplitude (\ref{4.17})
gives
\begin{equation}\label{4.19}
\Amplituda=
\frac{4\pi}{k}\sum_{\gamma}\mathrm{e}^{\mathrm{i}\delta_{\gamma}(E)}
\sin\delta_{\gamma}(E)
\mathcal{Y}_{\gamma}(\wektor{k}_{f})\mathcal{Y}_{\gamma}^{*}(\wektor{k}_{i}).
\end{equation}
As already mentioned the above result was obtained
\textit{without} prior construction of the scattering matrix. It
is also necessary to emphasize that the method used gives formula
for the scattering amplitude which has a similar form to that
obtained for potentials with spherical symmetry $V(r)$
\cite{Schiff}. For such potentials the functions
$\mathcal{Y}_{\gamma}(\wektor{k})$ reduce to the spherical
harmonics $Y_{lm}(\hat{\wektor{k}})$.
Subsequently, combining relation (\ref{4.19}) with (\ref{2.7}) we
have
\begin{equation}\label{4.20}
\sigma(\wektor{k}_{i})=\frac{16\pi^{2}}{k^{2}}\sum_{\gamma}
\sin^{2}\delta_{\gamma}(E)\mathcal{Y}_{\gamma}(\wektor{k}_{i})
\mathcal{Y}^{*}_{\gamma}(\wektor{k}_{i}).
\end{equation}
Finally, after substitution of Eq. (\ref{4.19}) to Eq. (\ref{2.8})
one obtains
\begin{equation}\label{4.21}
\sigma_{t}(E)=\frac{4\pi}{k^{2}}\sum_{\gamma}\sin^{2}\delta_{\gamma}(E).
\end{equation}
Thus, we have arrived at the well-known formula for the averaged
total cross section.
\section{Example}
\setcounter{equation}{0} \noindent\indent To illustrate the
method, let us consider scattering from a pair of identical
spheres, of radii $R$. The symmetry of this target allows us to
locate the origin of a coordinate system in the midpoint of the
interval joining the centers of spheres. Thus, we may choose the
spheres to be centered at points $\wektor{r}=\pm\wektor{\varrho}$.
However, due to the assumption of non-locality we may simulate
this collision process by potential
\begin{equation}\label{3.3.4}
V(\wektor{r},\wektor{r}')=\omega \left[
v_{+}(\wektor{r})v_{+}^{*}(\wektor{r}')
+v_{-}(\wektor{r})v_{-}^{*}(\wektor{r}') \right]
\end{equation}
with
\begin{equation}\label{3.3.3}
v_{\pm}(\wektor{r})=\frac{1}{\sqrt{4\pi}}
\frac{\delta(|\wektor{r}\pm\wektor{\varrho}|-R)} {R^{2}}.
\end{equation}
It should be noticed that potential defined by the Eqs.
(\ref{3.3.4}) and (\ref{3.3.3}) is the special case of that
proposed recently by de Prunel\'e \cite{Prunele} (see also
\cite{Prunele1}). As one can notice, for further simplicity the
strengths of both spheres were taken to be equal and have value
$\omega$. Therefore, the matrix $\Omega$ defined by Eq.
($\ref{3.7}$) may be rewritten as
\begin{equation}\label{3.3.5}
\mac{\Omega}=\begin{pmatrix}
\omega & 0\\
0 & \omega
\end{pmatrix} =\omega\mac{I}_{2},
\end{equation}
where $\mathsf{I}_{2}$ is the $2\times 2$ identity matrix.
Moreover, straightforward integrations in Eq. (\ref{elG}) with the
aid of the expansion
\begin{eqnarray}
&&\hspace{-0.8cm}\frac{\mathrm{e}^{\mathrm{i}k|\wektor{r}-
\wektor{r}'|}} {|\wektor{r}- \wektor{r}'|}= 4\pi \mathrm{i}k
\sum_{l=0}^{\infty}\sum_{m=-l}^{l}h_{l}^{(+)}(kr_{>})j_{l}(kr_{<})
Y_{lm}(\wektor{r})Y^{*}_{lm}(\wektor{r}')\nonumber\\
&&\hspace{5cm}(r_{>}=\mathrm{max}\{r,r'\},
r_{<}=\mathrm{min}\{r,r'\}).
\end{eqnarray}
gives the matrix $\mathsf{G}$ in the form
\begin{equation}\label{3.3.6}
\mac{G}=\mathrm{i}\eta k j_{0}(kR)
\begin{pmatrix}
h_{0}^{(+)}(kR) & j_{0}(kR) h_{0}^{(+)}(2k\varrho)\\ j_{0}(kR)
h_{0}^{(+)}(2k\varrho) & h_{0}^{(+)}(kR)
\end{pmatrix},
\end{equation}
where $\eta=2m/\hbar^{2}$, $j_{l}(z)$ stand for spherical Bessel
functions and $h_{l}^{(+)}(z)$ stand for Hankel functions of the
first kind. Then, utilizing Eqs. (\ref{4.2}) and (\ref{4.3}) one
finds
\begin{equation}\label{3.3.7}
\mac{A}=
\begin{pmatrix}
\omega^{-1}-\eta k j_{0}(kR)y_{0}(kR) & -\eta k j_{0}^{2}(kR)
y_{0}(2k\varrho)\\ -\eta k j_{0}^{2}(kR) y_{0}(2k\varrho) &
\omega^{-1}-\eta k j_{0}(kR)y_{0}(kR)
\end{pmatrix}
\end{equation}
and
\begin{equation}\label{3.3.8}
\mac{B}= \eta k j_{0}^{2}(kR)
\begin{pmatrix}
1 & j_{0}(2k\varrho)\\ j_{0}(2k\varrho) & 1
\end{pmatrix}
\end{equation}
with $y_{l}(z)$ denoting spherical Neumann functions. Note that,
in general
\begin{equation}\label{}
j_{0}(z)=\frac{\sin z}{z}, \qquad y_{0}(z)=-\frac{\cos
z}{z},\qquad
h_{0}^{(+)}(z)=-\mathrm{i}\frac{\mathrm{e}^{\mathrm{i}z}}{z}.
\end{equation}
As one can see the eigenvalue problem (\ref{4.6}) reduces to
\begin{equation}\label{3.3.9}
\mac{A}X_{\pm}(E)=\lambda_{\pm}\mac{B}X_{\pm}(E),
\end{equation}
where its eigenvalues $\lambda_{\pm}(E)$ are
\begin{equation}\label{3.3.10}
\lambda_{\pm}(E)= \frac{\omega^{-1}- \eta k j_{0}(kR) \left[
y_{0}(kR)\pm j_{0}(kR)y_{0}(2k\varrho) \right]} {\eta k
j_{0}^{2}(kR) [1\pm j_{0}(2k\varrho) ]}
\end{equation}
and respective eigenvectors
\begin{equation}\label{3.3.11}
X_{\pm}(E)= \left\{ 2\eta k j_{0}^{2}(kR) [ 1\pm j_{0}(2k\varrho)]
\right\}^{-1/2}\binom{1}{\pm1}.
\end{equation}
Since for arbitrary $\wektor{k}$:
\begin{equation}\label{3.3.12}
\big<\wektor{v}|\wektor{k}\big>=\left(
\begin{array}{c}
\big<v_{+}|\wektor{k}\big>\\*[0.5ex] \big<v_{-}|\wektor{k}\big>
\end{array}\right)= \sqrt{4\pi}\,j_{0}(kR)
\left(
\begin{array}{c}
\mathrm{e}^{-\mathrm{i}\wektor{k}\cdot\wektor{\varrho}}\\*[0.5ex]
\mathrm{e}^{\mathrm{i}\wektor{k}\cdot\wektor{\varrho}}
\end{array}\right)
\end{equation}
and by virtue of Eqs. (\ref{Yfun}) and (\ref{3.3.11}) the
eigenchannel harmonics may be expressed as
\begin{equation}\label{3.3.13}
\mathcal{Y}_{\pm}(\wektor{k})=\frac{1}{2\sqrt{2\pi}}\, [ 1\pm
j_{0}(2k\varrho) ]^{-1/2} \left(
\mathrm{e}^{-\mathrm{i}\wektor{k}\cdot\wektor{\varrho}}\pm
\mathrm{e}^{\mathrm{i}\wektor{k}\cdot\wektor{\varrho}} \right).
\end{equation}
and therefore, after substitution of Eqs. (\ref{3.3.10}) and
(\ref{3.3.13}) to Eq. (\ref{4.17}) we infer that the scattering
amplitude is of the form
\begin{eqnarray}\label{3.3.14}
&&\hspace{-0.7cm}\Amplituda=-2\eta j_{0}^{2}(kR)\left[
\frac{\cos(\wektor{k}_{f}\cdot\wektor{\varrho})\cos(\wektor{k}_{i}\cdot\wektor{\varrho})}
{\omega^{-1}+ \mathrm{i}\eta k j_{0}(kR)[h_{0}^{(+)}(kR)
+j_{0}(kR) h_{0}^{(+)}(2k\varrho)]}\right.\nonumber\\
&&\qquad\qquad\quad\left.+
\frac{\sin(\wektor{k}_{f}\cdot\wektor{\varrho})\sin(\wektor{k}_{i}\cdot\wektor{\varrho})}
{\omega^{-1}+ \mathrm{i}\eta k j_{0}(kR)[h_{0}^{(+)}(kR) -
j_{0}(kR) h_{0}^{(+)}(2k\varrho)]}\right].\nonumber\\
\end{eqnarray}
After application of the above to Eq. (\ref{2.7}) and integration
over all directions of scattering $\wektor{k}_{f}$ we infer
\begin{eqnarray}\label{3.3.15}
&&\hspace{-0.7cm}\sigma(\wektor{k}_{i})=8\eta^{2}j_{0}^{4}(kR)
\nonumber\\
&&\hspace{-0.8cm}\times \left[ \frac{[1+j_{0}(2k\varrho)]
\cos^{2}(\wektor{k}_{i}\cdot\wektor{\varrho})} { \left\{
\omega^{-1}-\eta k j_{0}(kR) \left[
y_{0}(kR)+j_{0}(kR)y_{0}(2k\varrho) \right]
\right\}^{2}+\eta^{2}k^{2} j_{0}^{4}(kR)[1+j_{0}(2k\varrho)]^{2}
}\right.\nonumber\\
&&\hspace{-0.8cm}+\left. \frac{ [1-j_{0}(2k\varrho)]
\sin^{2}(\wektor{k}_{i}\cdot\wektor{\varrho})} { \left\{
\omega^{-1}-\eta k j_{0}(kR) \left[
y_{0}(kR)-j_{0}(kR)y_{0}(2k\varrho) \right]
\right\}^{2}+\eta^{2}k^{2}j_{0}^{4}(kR) \left[ 1-j_{0}(2k\varrho)
\right]^{2}}
\right].\nonumber\\
\end{eqnarray}
Then averaging Eq. (\ref{3.3.15}) over all directions of incidence
$\wektor{k}_{i}$ and using the fact that
\begin{equation}\label{3.3.16}
\calkapow{k}_{i}\:\cos^{2}(\wektor{k}_{i}\cdot\wektor{\varrho})=2\pi
[1+j_{0}(2k\varrho)]
\end{equation}
and
\begin{equation}\label{3.3.17}
\calkapow{k}_{i}\:\sin^{2}(\wektor{k}_{i}\cdot\wektor{\varrho})=2\pi
[1-j_{0}(2k\varrho)],
\end{equation}
we obtain the averaged total cross section
\begin{eqnarray}\label{3.3.18}
&&\hspace{-0.6cm}\sigma_{t}(E)=4\pi\eta^{2} j_{0}^{4}(kR)
\nonumber\\
&&\hspace{-0.6cm}\times \left[ \frac{[1+j_{0}(2k\varrho)]^{2}} {
\left\{ \omega^{-1}-\eta k j_{0}(kR)
[y_{0}(kR)+j_{0}(kR)y_{0}(2k\varrho)] \right\}^{2}+\eta^{2}k^{2}
j_{0}^{4}(kR)[1+j_{0}(2k\varrho)]^{2}
}\right.\nonumber\\
&&\hspace{-0.6cm}+\left. \frac{[1-j_{0}(2k\varrho)]^{2}} { \left\{
\omega^{-1}-\eta k j_{0}(kR) [y_{0}(kR)-j_{0}(kR)y_{0}(2k\varrho)]
\right\}^{2}+\eta^{2}k^{2}j_{0}^{4}(kR)[1-j_{0}(2k\varrho)]^{2}}
\right].\nonumber\\
\end{eqnarray}
Note that the results (\ref{3.3.17}) and (\ref{3.3.18}) may be
equivalently obtained after application of Eqs. (\ref{3.3.10}) and
(\ref{3.3.13}), respectively, to Eqs. (\ref{4.20}) and
(\ref{4.21}).

\noindent\textbf{Acknowledgements} I am grateful to R.~Szmytkowski
for very useful discussions, suggestions and commenting on the
manuscript. Discussions with P.~Horodecki are also acknowledged.

\appendix
\section{Positive semidefiniteness of the matrix $\mathsf{B}$}
\label{appB} \setcounter{equation}{0}

The proof is due to R.~Szmytkowski \cite{Szmytkowski4}. Below we
show that the matrix $\mac{B}$ given by Eq. (\ref{4.3}) is such
that, the inequality
\begin{equation}\label{A.1}
X_{\gamma}^{\dagger}(E)\mac{B}X_{\gamma}(E)\ge 0
\end{equation}
is satisfied. Since
\begin{equation}\label{A.2}
\calkapow{k}\,\mathrm{e}^{\mathrm{i}\wektor{k}\cdot
(\wektor{r}-\wektor{r}')}=4\pi\frac{\sin
k|\wektor{r}-\wektor{r}'|}{|\wektor{r}-\wektor{r}'|}
\end{equation}
we may rewrite Eq. (\ref{elB}) in the form
\begin{equation}\label{A.3}
[\mac{B}]_{\nu\mu}=\frac{mk}{8\pi^{2}\hbar^{2}}\calkapow{k}\calkaob{r}\,v_{\nu}^{*}(\wektor{r})
\mathrm{e}^{\mathrm{i}\wektor{k}\cdot\wektor{r}}\calkaob{r}'\,v_{\mu}(\wektor{r}')
\mathrm{e}^{-\mathrm{i}\wektor{k}\cdot\wektor{r}'},
\end{equation}
which after application to Eq. (\ref{A.1}) yields
\begin{equation}\label{A.4}
X_{\gamma}^{\dagger}(E)\mac{B}X_{\gamma}(E)=\frac{mk}{8\pi^{2}\hbar^{2}}\calkapow{k}
\left|\sum_{\nu}X_{\gamma\nu}^{*}(E)\calkaob{r}\,
v_{\nu}^{*}(\wektor{r})\mathrm{e}^{\mathrm{i}\wektor{k}\cdot\wektor{r}}\right|^{2}\ge
0.
\end{equation}
Obviously, the above statement finishes the proof.
\section{Proof of orthonormality relation (\ref{orthrel})}
\label{appA} \setcounter{equation}{0}
Substituting Eq. (\ref{Yfun}) to Eq. (\ref{orthrel}) and
reorganizing terms one finds
\begin{eqnarray}\label{B.1}
&&\hspace{-0.5cm}\calkapow{k}\;\mathcal{Y}_{\gamma'}^{*}(\wektor{k})\mathcal{Y}_{\gamma}(\wektor{k})=
\frac{m k}{8\pi^{2}\hbar^{2}}
\nonumber\\
&&\times\sum_{\nu\mu}X_{\gamma'\nu}^{*}(E)
\calkaob{r}\,v^{*}_{\nu}(\wektor{r})\calkaob{r}'\,v_{\mu}(\wektor{r}')
\calkapow{k}\,\mathrm{e}^{\mathrm{i}\wektor{k}\cdot
(\wektor{r}-\wektor{r}')}X_{\gamma\mu}(E).\nonumber\\
\end{eqnarray}
In virtue of Eq. (\ref{A.2}), Eq. (\ref{B.1}) may be rewritten in
the form
\begin{eqnarray}\label{B.2}
&&\hspace{-1cm}\calkapow{k}\;\mathcal{Y}_{\gamma'}^{*}(\wektor{k})\mathcal{Y}_{\gamma}(\wektor{k})=
\frac{m}{2\pi\hbar^{2}}\nonumber\\
&&\times\sum_{\nu\mu}X_{\gamma'\nu}^{*}(E)\left[\calkaob{r}\calkaob{r}'\,v^{*}_{\nu}(\wektor{r})
\frac{\sin
k|\wektor{r}-\wektor{r}'|}{|\wektor{r}-\wektor{r}'|}v_{\mu}(\wektor{r}')\right]X_{\gamma\mu}(E).\nonumber\\
\end{eqnarray}
As one can notice with the aid of Eq. (\ref{elB}), the expression
in square brackets is proportional to an element of the matrix
$\mac{B}$. Hence, we arrive at
\begin{equation}\label{B.3}
\calkapow{k}\;\mathcal{Y}_{\gamma'}^{*}(\wektor{k})\mathcal{Y}_{\gamma}(\wektor{k})=
X_{\gamma'}^{\dagger}(E)\mac{B}X_{\gamma}(E),
\end{equation}
which after comparison with Eq. (\ref{4.8}) completes the proof.

\end{document}